\documentclass[]{eptcs}

\usepackage[font=small,labelfont=it]{caption}
\input{header.extra}

\begin{document}

\title{\bf Surface proofs for nonsymmetric linear logic\\(extended abstract)}


\title{Surface Proofs for Nonsymmetric Linear Logic}
\author{Lawrence Dunn
\institute{North Florida Community College, USA}
\email{dunnl@nfcc.edu}
\and
Jamie Vicary
\institute{University of Oxford, UK}
\email{jamie.vicary@cs.ox.ac.uk}
}
\def\titlerunning{Surface Proofs for Nonsymmetric Linear Logic}
\def\authorrunning{L. Dunn and J. Vicary}


\date{January 20, 2016}
\maketitle

\begin{abstract}
We show that a proof in multiplicative linear logic can be represented as a decorated surface, such that two proofs are logically equivalent just when their  surfaces are geometrically equivalent. This is an extended abstract for \href{http://arxiv.org/abs/1601.05372}{arXiv:1601.05372}.
\end{abstract}

\section{Introduction}

Multiplicative linear logic~\cite{sep-ll, girard} is a  formal calculus for reasoning about resources, which is similar to traditional logic, except that resources cannot be duplicated or neglected in the way that propositions can.

A central problem in logic is determining when two proofs should be considered equivalent. In this paper, we describe a scheme for interpreting proofs in multiplicative linear logic as \textit{geometrical surfaces} embedded in 3d space. We define two surfaces as \textit{equivalent} just when one can be deformed into the other, in sense we make precise. Our main theorem then reads as follows.

\begin{customthm}{\ref{thm:maintheorem}}\em
Two sequent proofs in nonsymmetric multiplicative linear logic have equal interpretations in the free $*$-autonomous category just when their surfaces are equivalent.
\end{customthm}

\noindent
The theory of $*$-autonomous categories~\cite{mellies-categorical, seely-linearlogic} is a standard mathematical model for linear logic, so this theorem says that the notion of proof equality provided by the surface calculus agrees with the standard one.

Our surfaces are similar in spirit to {proof nets}~\cite{blute-coherence, girard}; however, we argue that our scheme has several advantages. In particular, correctness is local; any well-typed composite produces a valid proof-theoretic object, with no global property, such as the long-trip criterion, to be verified. Also, our notion of equivalence is broad, establishing some proof equivalences in fewer steps than for proof nets; sometimes in just a single step. Our scheme also certainly has disadvantages: in particular, we do not present a decision procedure for equivalence of our 3d diagrams, although we expect such a procedure could be described. Despite these differences, the formalisms are intimately connected, in the following way: the proof net is the 2d \textit{projection} of the 3d surface geometry. From this perspective, we can make sense of some of the features of proof nets: the long-trip criterion can be interpreted as a non-local check that the 2d shadow is consistent with a valid 3d geometry, and the thinning link decorations indicate the depth at which a unit is attached in the 3d geometry.

The underlying technical contribution, which we do not describe in this extended abstract,  is a direct algebraic proof of the  coherence theorem for Frobenius pseudomonoids, which shows that all diagrams of a certain sort commute. This is more appropriate for logical purposes, and more flexible, than an existing topological proof arising from Morse theory~\cite{KerlerLyubashenko, Lauda:2005}.

The authors are grateful to Samson Abramsky, Nick Gurski, Sam Staton and the anonymous reviewers for useful comments. 3d graphics have been written in Ti\textit{k}Z, and 2d graphics have been produced by the proof assistant Globular~\cite{globular}.

\subsection{Related work}
\label{sec:related}

It is well-recognized that ideas from topology are relevant for linear logic. The original proof nets of Girard~\cite{girard} are topological objects,  and Melli\`es has shown how the topology of ribbons gives a decision procedure for correctness of proof nets~\cite{mellies-ribbon}. Proof nets allow reasoning about proofs with units, but the formalism is complex, requiring a system of thinning links with moving connections~\cite{blute-coherence}. Hughes~\cite{hughes-simpleproof} gives a variant of proof nets which works well with units, and which has good compositional properties, but which still requires a long-trip criterion, has non-local jumps, and requires successive individual re-wirings. Our approach has a local flavour which is shared by the deep-inference model of proof analysis~\cite{guglielmi-deepinference} and the access to monoidal coherence that it allows~\cite{hughes-deepinference}; however, the coherence property we make use of is strictly more powerful, as it operates in a fragment that combines the $\otimes$ and $\parr$ connectives. We note also the work of Slavnov~\cite{slavnov} on linear logic and surfaces, which involves some similar ideas to the present article, but is technically quite unrelated.

\section{Surface calculus}
\label{sec:surfacecalculus}

In this section we develop the 2d string diagram calculus for sequents, and the 3d surface calculus for proofs. We show how to translate a sequent calculus proof into the surface calculus, and we define the equivalence relation on surfaces.

\subsection{The 2d calculus}
\label{sec:2dcalculus}
\tikzset{extrascale/.style={scale=0.8}}

The 2d calculus, which we will use to represent individual sequents, is the Joyal-Street calculus for monoidal categories~\cite{Joyal_1991}, directed from left to right. We use the standard 2-sided sequent calculus for nonsymmetric multiplicative linear logic with units~\cite{blute-coherence}: our sequents are pairs $\Gamma \vdash \Delta$, where $\Gamma$ and $\Delta$ are ordered lists (separated with ``,'') of expressions in the following grammar, where $V=\{A,B,C,\ldots\}$ is a set of atomic variables:
\[
S::= I \,\,|\,\, \bot \,\,|\,\, V \,\,|\,\, S \otimes S \,\,|\,\, S \parr S \,\,|\,\, S^* \,\,|\,\, \ls S
\]
We have left and right negation, and isomorphisms $\ls(S^*) \simeq S \simeq (\ls S)^*$ are a native part of the calculus (see \autoref{ex:tripleunit}); for simplicity, we suppress them at the syntactic level.

Our graphical language for sequents is as follows. Atomic variables are represented as black dots, pointing in different directions depending on their side of the sequent:
\begin{calign}
\nonumber
\begin{tz}
\node (1) [inner sep=0pt] at (0,0) {\tikzpng[scale=3, rotate=-90, extrascale]{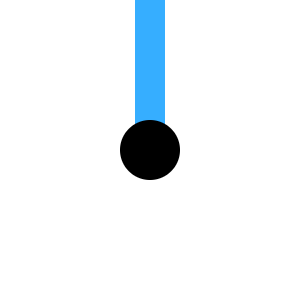}};
\draw [lightblue, dotted, line width=1.5pt] (1.east) to +(0.3,0);
\node [left] at (0,0) {$A$};
\end{tz}
&
\begin{tz}
\node (1) [inner sep=0pt] at (0,0) {\tikzpng[scale=3, rotate=90, extrascale]{blackdot}};
\draw [lightblue, dotted, line width=1.5pt] (1.west) to +(-0.3,0);
\node [right] at (0,0) {$A$};
\end{tz}
\\\nonumber
A \vdash \cdots & \cdots \vdash A
\end{calign}
These diagrams have nonempty boundary, in the sense that not all the ends are terminated by nodes.

The two sides of a sequent are represented graphically by trees, which are drawn connected together at their roots. The basic connective ``,'' is denoted as a blue vertex with zero or more branches to the left or right, as follows:
\tikzset{dot/.style={circle, draw=none, fill=black, inner sep=1.0pt}}
\begin{calign}
\nonumber
\begin{tz}[scale=0.76, extrascale]
\node [anchor=south west, inner sep=0pt] at (0,0) {\tikzpng[scale=3, rotate=90, extrascale]{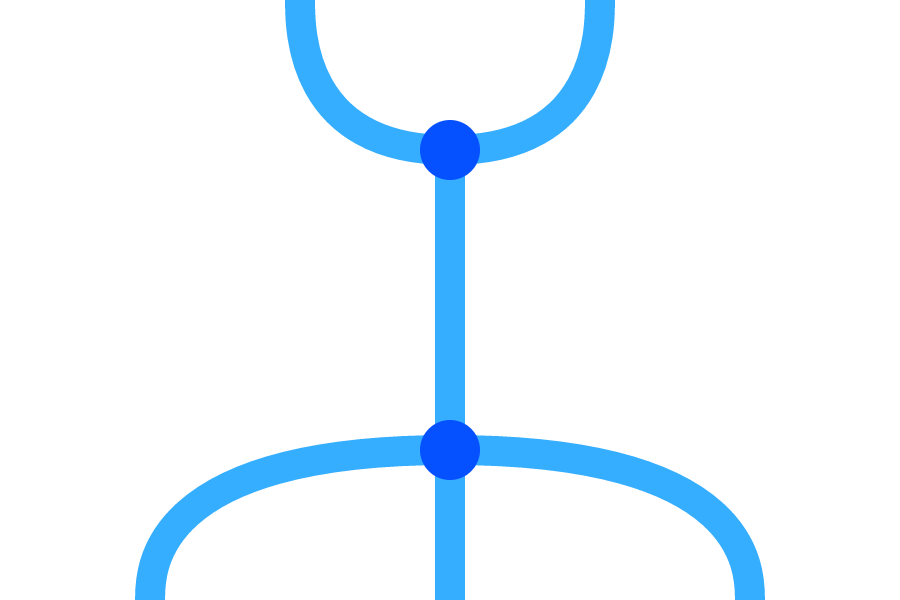}};
\node [dot] at (0,1) {};
\node [left] at (0,1) {$B$};
\node [dot] at (0,2) {};
\node [left] at (0,2) {$A$};
\node [dot] at (2,2.5) {};
\node [right] at (2,2.5) {$C$};
\node [dot] at (2,1.5) {};
\node [right] at (2,1.5) {$D$};
\node [dot] at (2,0.5) {};
\node [right] at (2,0.5) {$E$};
\end{tz}
&
\begin{tz}[scale=0.76, extrascale]
\node [anchor=south west, inner sep=0pt] at (0,0) {\tikzpng[scale=3, rotate=90, extrascale]{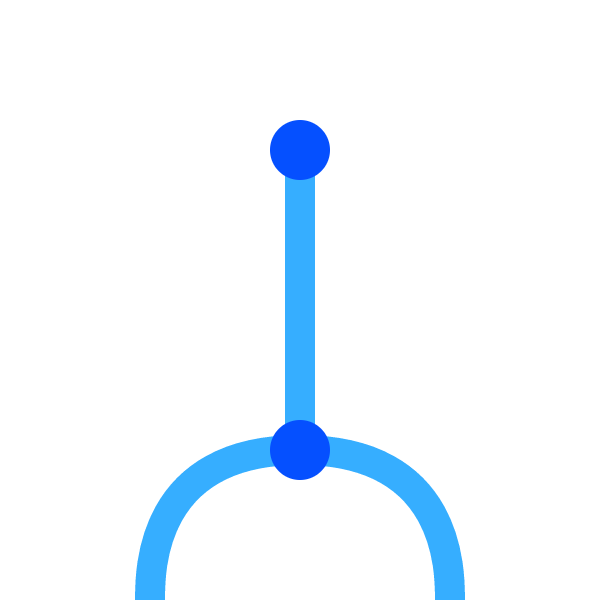}};
\node [dot] at (2,1.5) {};
\node [right] at (2,1.5) {$A$};
\node [dot] at (2,0.5) {};
\node [right] at (2,0.5) {$B$};
\end{tz}
&
\begin{tz}[scale=0.76, extrascale]
\node [anchor=south west, inner sep=0pt] at (0,0) {\tikzpng[scale=3, rotate=-90, extrascale]{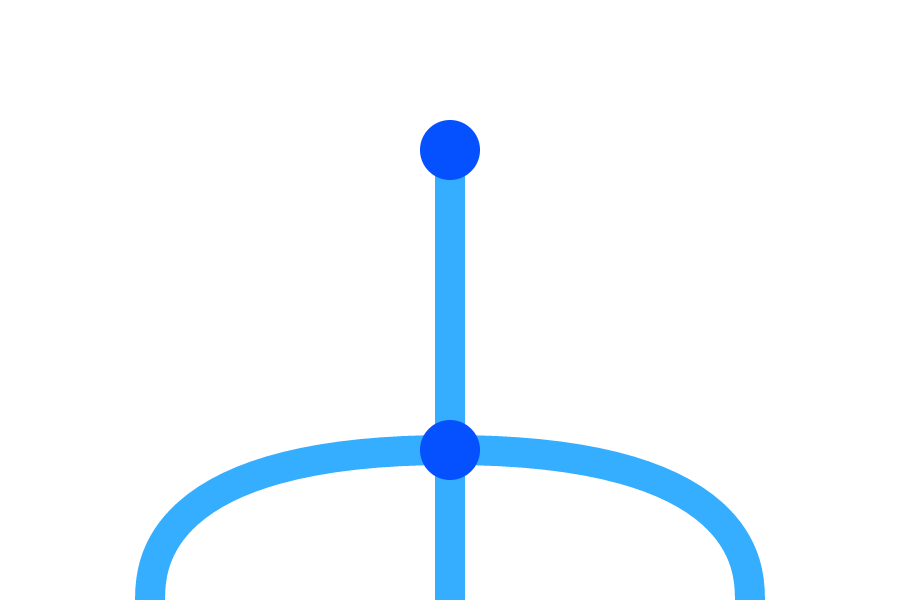}};
\node [dot] at (0,2.5) {};
\node [left] at (0,2.5) {$A$};
\node [dot] at (0,1.5) {};
\node [left] at (0,1.5) {$B$};
\node [dot] at (0,0.5) {};
\node [left] at (0,0.5) {$C$};
\end{tz}
\\
\nonumber
A,B \vdash C, D, E
&
\vdash A,B
&
A, B, C \vdash
\end{calign}
The connectives $\otimes$ and $\parr$, which are always binary, are drawn in blue on their natural side (left for $\otimes$, right for $\parr$), and in red on the other side, as we show with the following examples:
\begin{calign}
\nonumber
\begin{tz}[scale=0.76, extrascale]
\node [anchor=south west, inner sep=0pt] at (0,0) {\tikzpng[scale=3, rotate=-90, extrascale]{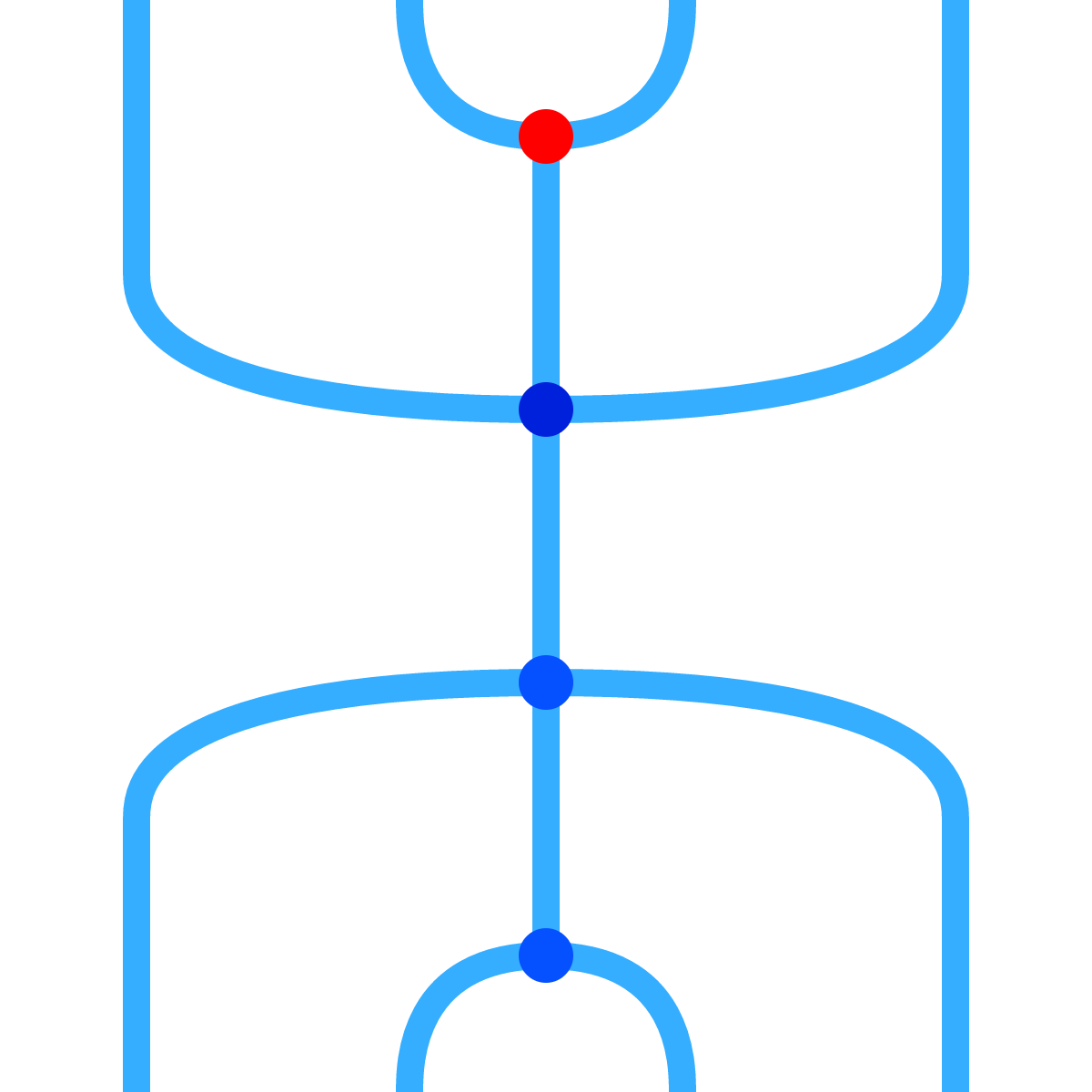}};
\node [dot] at (0,3.5) {};
\node [left] at (0,3.5) {$A$};
\node [dot] at (0,2.5) {};
\node [left] at (0,2.5) {$B$};
\node [dot] at (0,1.5) {};
\node [left] at (0,1.5) {$C$};
\node [dot] at (0,0.5) {};
\node [left] at (0,0.5) {$D$};
\node [dot] at (4,3.5) {};
\node [right] at (4,3.5) {$E$};
\node [dot] at (4,2.5) {};
\node [right] at (4,2.5) {$F$};
\node [dot] at (4,1.5) {};
\node [right] at (4,1.5) {$G$};
\node [dot] at (4,0.5) {};
\node [right] at (4,0.5) {$H$};
\end{tz}
&
\begin{tz}[scale=0.76, extrascale]
\node [anchor=south west, inner sep=0pt] at (0,0) {\tikzpng[scale=3, rotate=-90, extrascale]{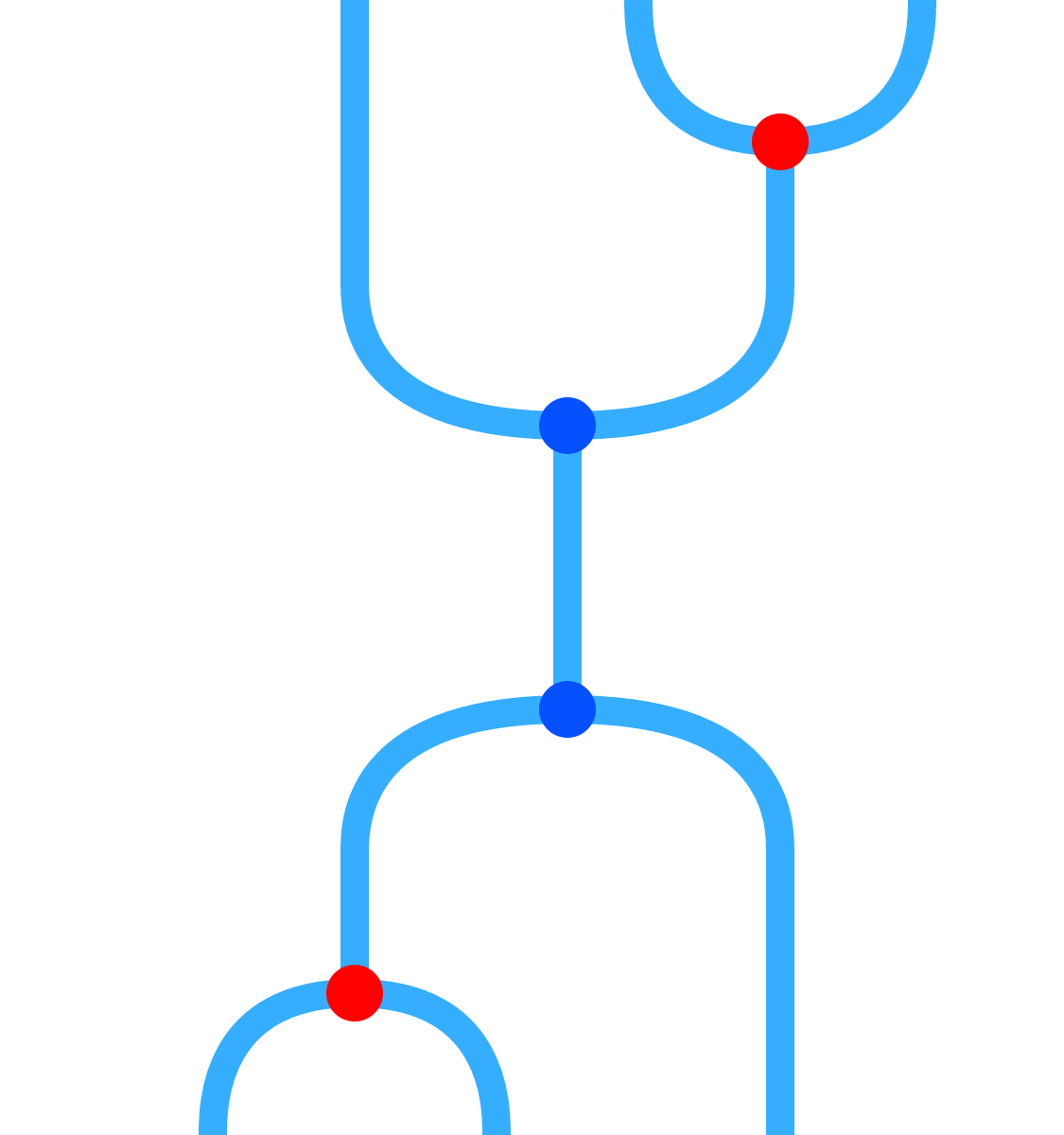}};
\node [dot] at (0,3) {};
\node [left] at (0,3) {$A$};
\node [dot] at (0,2) {};
\node [left] at (0,2) {$B$};
\node [dot] at (0,1) {};
\node [left] at (0,1) {$C$};
\node [dot] at (4,2.5) {};
\node [right] at (4,2.5) {$D$};
\node [dot] at (4,1.5) {};
\node [right] at (4,1.5) {$E$};
\node [dot] at (4,0.5) {};
\node [right] at (4,0.5) {$F$};
\end{tz}
\\*\nonumber
A, (B \otimes C), D \vdash E, (F \otimes G), H
&
(A \parr B) , C \vdash D \parr (E \otimes F)
\end{calign}
Note that a blue dot with a binary branching is therefore an overloaded notation; this is a deliberate feature.

The units $I$ and $\bot$ are represented by blue dots on their natural side (left for $I$, right for $\bot$), and red dots on the other side, as shown:
\begin{calign}
\nonumber
\begin{tz}[scale=0.76, extrascale]
\node [anchor=south west, inner sep=0pt] at (0,0) {\tikzpng[scale=3, rotate=-90, extrascale]{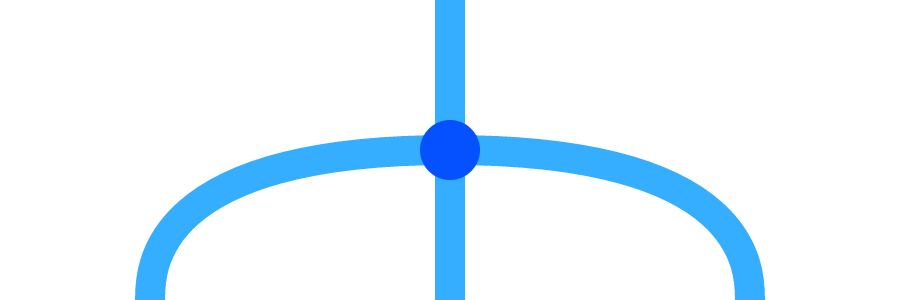}};
\node [dot] at (0,2.5) {};
\node [left] at (0,2.5) {$A$};
\node [dot, fill=darkblue] at (0,1.5) {};
\node [left] at (0,1.5) {};
\node [dot] at (0,0.5) {};
\node [left] at (0,0.5) {$B$};
\node [dot] at (1,1.5) {};
\node [right] at (1,1.5) {$C$};
\end{tz}
&
\begin{tz}[scale=0.76, extrascale]
\node [anchor=south west, inner sep=0pt] at (0,0) {\tikzpng[scale=3, rotate=90, extrascale]{comma}};
\node [dot, fill=red] at (0,1) {};
\node [left] at (0,1) {};
\node [dot] at (0,2) {};
\node [left] at (0,2) {$A$};
\node [dot] at (2,2.5) {};
\node [right] at (2,2.5) {$B$};
\node [dot, fill=red] at (2,1.5) {};
\node [right] at (2,1.5) {};
\node [dot, fill=darkblue] at (2,0.5) {};
\node [right] at (2,0.5) {};
\end{tz}
\\
\nonumber
A,I,B \vdash C
&
A, \bot \vdash B, I, \bot
\end{calign}
We represent $(-)^*$ as turning right by a half-turn, and ${}^* \hspace{-0.5pt} (-)$ as turning left by a half-turn, as shown:
\begin{calign}
\nonumber
\begin{tz}[scale=0.76, extrascale]
\node [anchor=south west, inner sep=0pt] at (0,0) {\tikzpng[scale=3, rotate=90, extrascale]{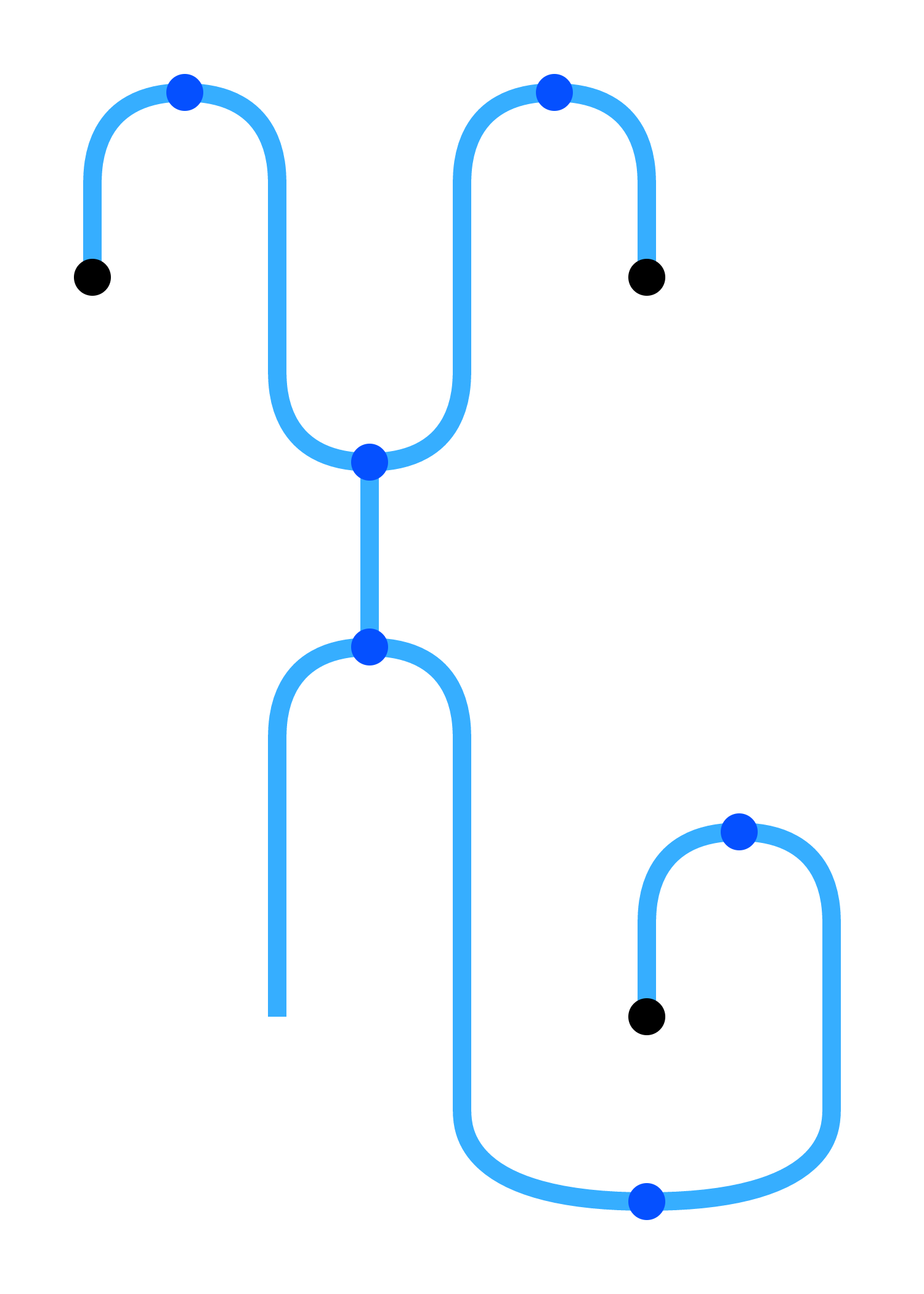}};
\node [right] at (1.5,3.5) {$A$};
\node [right] at (1.5,0.5) {$B$};
\node [dot] at (5.5,1.5) {};
\node [right] at (5.5,3.5) {$C$};
\node [right] at (5.5,1.5) {$D$};
\end{tz}
\\
\nonumber
A^* , \ls B \vdash \ls \,\ls \,C, D
\end{calign}
Diagrams built from  sequents in this way are of a simple kind; as graphs, they are all acyclic and connected. In general we can allow arbitrary well-typed composites of these components; such diagrams represent 1\-morphisms in the monoidal bicategory \free {\F^*}, described in the full paper.

\subsection{The 3d calculus}

Diagrams in the 3d calculus are surfaces embedded in $\R^3$. Formally they are expressions in the graphical calculus for Gray categories, which is by now well-developed~\cite{barrett-graydiagrams, bartlett-wire, hummon-thesis, CSPthesis}. However, the 3d calculus is quite intuitive, and we take advantage of this to introduce it in an informal way.

Diagrams consist of \textit{sheets}, bounded on the left and right by \textit{edges}, which are bounded above and below by \emph{vertices}. (Sheets can also be bounded by the sides of the diagram, and edges can also be bounded by the top or bottom of the diagram.) Diagrams are immersed in 3d space, meaning that sheets can exist in front or behind other sheets, and wires on sheets of different depths can cross; however, components never intersect. Here is an example:
\begin{equation*}
\tikzset{blob/.style={draw, circle, fill=black, inner sep=2pt}}
\tikzset{every picture/.style={scale=0.7}}
\begin{tz}[scale=1.5]
\draw [surface] (\xoff,\yoff) node (1) {} to (3+\xoff,\yoff) node (2) {} to (3+\xoff,3+\yoff) node (3) {} to (\xoff,3+\yoff) node (4) {} to (1.center);
\draw [black] (2,\yoff) to [out=up, in=down] (1,1+\yoff) node (x) {} to (1,3+\yoff);
\node [blob] at (x.center) {};
\draw [surface] (-\xoff,-\yoff) node (5) {} to (3-\xoff,-\yoff) node (6) {} to (3-\xoff,3-\yoff) node (7) {} to (-\xoff,3-\yoff) node (8) {} to (5.center);
\draw [black] (1,-\yoff) to (1,\yoff) to [out=up, in=down] (2,1+\yoff) to (2,2) to [out=up, in=down] (2,3-\yoff);
\node [blob] at (2,2) {};
\end{tz}
\end{equation*}
Here we have front and back sheets, each containing an edge, which contains a vertex. Towards the bottom of the picture, the wires cross: this is called an \emph{interchanger}.

For our application to linear logic, we allow two types of vertex: \emph{coherent vertices} and \emph{adjunction vertices}.
\begin{itemize}
\item
\textbf{Coherent vertices.} Say that a 2d calculus diagram is \textit{simple} when it is connected and acyclic with nonempty boundary, and in the blue fragment of the calculus, not involving red nodes or black atomic variable nodes. Then any two simple diagrams can be connected by a coherent vertex, denoted as follows:
\begin{calign}
\label{eq:surfaceexample}
\begin{tz}[scale=1.5]
\draw [surface] (-\xoff,2) node (1) {} to (0.8,2) node (2) {} to [out=45, in=left, looseness=0.4] (3-\xoff,2+\yoff) node (3) {} to (3-\xoff,\yoff) node (4) {} to [out=left, in=45, in looseness=0.6] (2,0) node (5) {} to (1,0) node (6) {} to [out=145, in=right] (-\xoff, \yoff) node (7) {} to (1.center) to (2.center);
\draw [surface] (2.center) to (1.5,0.8) node (8) {} to (2.2,2-\yoff) node (9) {} to (2.center);
\draw [surface] (6.center) to (8.center) to (9.center) to [out=-165, in=right] (\xoff,2-2*\yoff) node (14) {} to (\xoff,-\yoff) node (13) {} to [out=right, in=-155] (6.center);
\draw [surface] (9.center) to (3+\xoff,2-\yoff) node (15) {} to (3+\xoff,-\yoff) node (16) {} to [out=left, in=-35] (5.center) to (8.center) to (9.center);
\end{tz}
&
\begin{tz}
\node [inner sep=0pt] (1) at (0,1.5) {\tikzpng[scale=2, rotate=90]{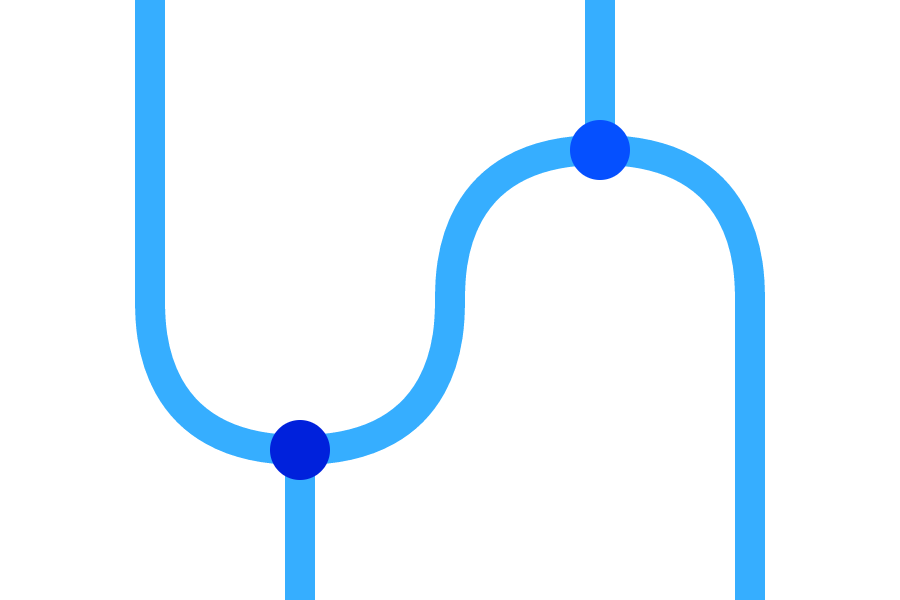}};
\node [inner sep=0pt] (2) at (0,0) {\tikzpng[scale=2, rotate=90]{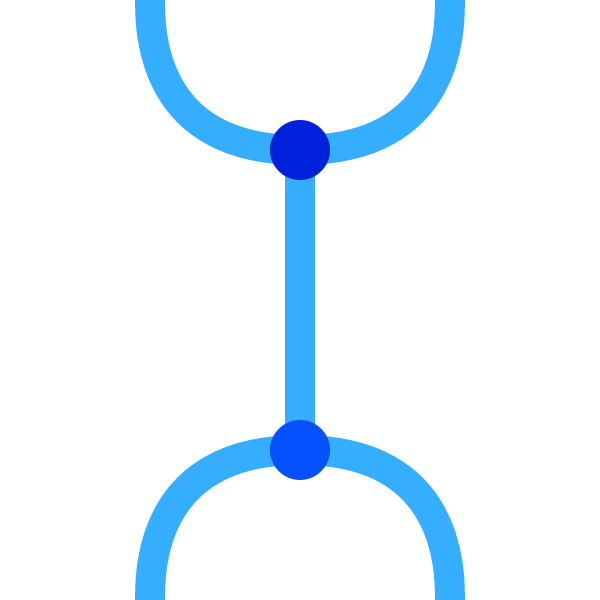}};
\draw [->, shorten <=-2pt, shorten >=-2pt] (1) to (2);
\end{tz}
\end{calign}
On the left we give the surface representation, and on the right we give the 2d calculus representation of the upper and lower boundaries. The coherent vertex is the point in the middle of the surface diagram where 4 edges meet.
\item
\textbf{Adjunction vertices.} Listed in \autoref{fig:3dgenerators}, these introduce and eliminate red and black edges in the surface calculus.
\end{itemize}
\input{3dgenerators.extra}
\input{3dequations.extra}

We now define equivalence in the graphical language, giving {intuitive} interpretations of each generating relation in italics. 
\begin{definition}
\label{def:surfaceequivalence}
Two surface diagrams are \textit{equivalent} when they are related by the least equivalence relation generated by the following:
\begin{itemize}
\item \textbf{Coherence.} Let $P,Q$ be surface diagrams built from  coherent vertices, with equal lower boundaries and equal upper boundaries, with all these boundaries being simple; then $P=Q$. \emph{(All acyclic equations of coherent vertices hold.)}
\item \textbf{Adjunction.} The equations listed in \autoref{fig:3dequations} hold. \emph{(Bent wires can be pulled straight.)}
\item \textbf{Isotopy.} The equations of a monoidal bicategory hold. \emph{(If two diagrams are ambient isotopic, they are equivalent.)}
\item \textbf{Locality.} Suppose surface diagrams $P,Q$ differ only with respect to subdiagrams $P',Q'$, with $P'=Q'$. Then $P=Q$. \emph{(Equivalence applies locally in the interior of a diagram.)}
\end{itemize}
\end{definition}

\noindent
It is a fair summary of this definition to say that two diagrams are equivalent just when one can be \textit{deformed} into the other. We emphasize that our contribution here is the \textbf{Coherence} axiom, which has not previously been noted\footnote{In the full version of this paper we give a direct combinatorial proof of the coherence property, which also follows from topological arguments due to Kerler and Lyubashenko~\cite{KerlerLyubashenko} as refined by Lauda~\cite{Lauda:2005}.}; the rest follows in principle from the work of Street~\cite{Street_2004}, although it has not to our knowledge been explicitly described in the literature, and its implications for linear logic unpacked. However, we note that it is the \textbf{Coherence} axiom that makes the notion of equivalence tractable.

Our presentation here is informal, but we emphasize that our definition of the surface calculus and its equivalence relation can be made completely precise in terms of the formal development of the full article: two diagrams are equivalent just when they are equal as 2\-morphisms in the monoidal bicategory $\free{\F ^*}$.

\subsection{Interpreting the sequent calculus}

We saw in \autoref{sec:2dcalculus} how individual sequents in multiplicative linear logic can be interpreted as 2d diagrams. We now see how proofs can be interpreted as 3d surface diagrams. We view these surfaces as directed from top to bottom, just like traditional sequent calculus proofs; so for a particular surface, its \textit{hypothesis} is the upper boundary, and its \textit{conclusion} is the lower boundary.

We use a basis for the sequent calculus with a symmetry between introduction and elimination for $\otimes$, $\parr$, $I$ and $\bot$; the rules $\otimes$\-R, $\parr$\-L, $I$\-R and $\bot$\-L are derivable (see \autoref{ex:additionalrules}.) Furthermore, we include only CUT rules with minimal overlapping contexts; the more general CUT rules are derivable using negation. (These two features account for the differences between our presentations and others in the literature~\cite{Abrusci:1991}.) The interpretation of AXIOM and CUT rules are given recursively in \autoref{fig:3dcompound}, with black wires standing for atomic variables and green wires standing for general variables; the interpretation of the remaining rules, which we call the \emph{core fragment} of the logic, is given in \autoref{fig:sequentscore}. We now state our main theorem.
\input{3dcompound.extra}
\input{sequents-core.extra}
\begin{theorem}
\label{thm:maintheorem}
Two sequent proofs in multiplicative linear logic have equal interpretations in the free $*$-autonomous category just when their surface diagrams are equivalent.
\end{theorem}

\noindent
It is interesting to analyze the different contributions to proof equivalence made by each part of \autoref{def:surfaceequivalence} of surface equivalence. \textbf{Coherence} tells us that any two proofs built in the core part of the logic given in \autoref{fig:sequentscore} are equal. \textbf{Adjunction} tells us that AXIOM and CUT cancel each other out, both for atomic and compound variables. \textbf{Isotopy} tells us that that `commutative conversion' is possible, where by exchanging heights of disconnected parts of the diagram, we exchange the order of separate sequent calculus proof steps. \textbf{Locality} tells us that we can apply our equations in the context of a larger proof, in the manner of deep inference~\cite{guglielmi-deepinference}.

We give a formal statement of coherence for the core part of the logic, since it is a result of independent interest. Note that this is not a theorem about the surface calculus, although its proof uses the surface calculus.
\begin{corollary}
If two sequent proofs in the core fragment of the logic given in \autoref{fig:sequentscore} have the same hypotheses and conclusion, then they are equal in the free $*$\-autonomous category.
\end{corollary}

\noindent
We comment on some interesting features of the translation between the sequent calculus and the surface calculus. The fundamental simplicity of the surface calculus is clear, from the minimality of the data in \autoref{fig:3dgenerators}, as compared to Figures~\ref{fig:3dcompound} and~\ref{fig:sequentscore}. Partly this is achieved by the greater degree of locality: for example, the cut rules for $\ls V$ and $V^*$ are both interpreted using the same surface generators, composed in different ways. But more significantly, the entire core fragment of the sequent calculus is interpreted in the \emph{trivial} part of the surface calculus, significantly reducing the bureaucracy of proof analysis, to use Girard's phrasing~\cite{girard}. To make the most of these advantages, we suggest that the surface calculus can  serve {directly} as a toolkit for logic, not just as a way to visualize sequent calculus proofs.

\section{Examples}
\label{sec:examples}

\input{timesRproof.extra}
\input{unitproof.extra}

In this Section, we look in detail at a number of examples: we derive the surface form of the missing $\otimes$\-R rule; we analyze equivalence of a proof involving units; and we investigate the classic triple-unit problem.

\begin{example}[Additional rules]
\label{ex:additionalrules}
Presentations of multiplicative linear logic usually include the rules $\otimes$-R, $\parr$-L, $I$-R, $\bot$-L, which are missing from \autoref{fig:3dcompound} and \autoref{fig:sequentscore}; however, they are derivable. We analyze $\otimes$-R in detail in \autoref{figure:tensorright}. On the left-hand side, we derive the rule in our chosen basis for the the sequent calculus. In the middle image, we interpret it in the surface calculus, using the rules we have described. In the third image, we simplify the surface calculus interpretation using the rules in \autoref{fig:3dequations}. From this simplified diagram, we see that it does not in fact involve the variables, the nontrivial generators being applied in the central part of diagram only. Elegant interpretations of the other 3 rules can be derived similarly.
\end{example}

\begin{example}[Triple-dual problem]
\label{ex:tripleunit}
Starting with the identity \mbox{$A \multimap X \to A \multimap X$}, we can uncurry on the left to obtain a morphism \mbox{$A \otimes (A \multimap X) \to X$}, and curry on the right to obtain a morphism $p_A : A \to X \multimapinv (A \multimap X)$; in a similar way, we can also define a morphism $q_A : A \to (X \multimapinv A) \multimap X$. Then the \textit{triple-dual problem}, originally due to Kelly and Mac Lane~\cite{Kelly_1979} and generalized here to the non-symmetric setting, is to determine whether the following equation holds:
\begin{equation}
\label{eq:tripleunit}
\begin{aligned}
\begin{tikzpicture}[xscale=1.5,yscale=2]
\node (1) at (0,0) {$X \multimapinv ((X \multimapinv A) \multimap X)$};
\node (2) at (4,0) {$X \multimapinv A$};
\node (3) at (4,-1) {$X \multimapinv ((X \multimapinv A) \multimap X)$};
\draw [->] (1) to node [above] {$X \multimapinv q_A$} (2);
\draw [->] (2) to node [right] {$p_{X \multimapinv A}$} (3);
\draw [->] (1) to node [below] {$\id$} (3);
\end{tikzpicture}
\end{aligned}
\end{equation}
\begin{figure}[t!]
$$
\hspace{-5cm}
\begin{aligned}
\begin{tikzpicture}
\path [use as bounding box] (4.5,-0.75) rectangle +(3.7,5);
\node (4) [right] at (4-\xoff+0.2,1.5) {$\tree[$\otimes$-R]{}{A, B \vdash \bot, A \otimes B}$};
\node (1) [anchor=west] at (0,4.05 -| 4.west) {$\tree[AXIOM]{}{A \vdash A}$};
\node (2) [anchor=west] at (0,3.2 -| 4.west) {$\tree[$\bot$-INT]{}{A \vdash \bot, A}$};
\node [anchor=west] at (5.2,2.55) {$\tree[AXIOM]{}{B \vdash B}$};
\end{tikzpicture}
\end{aligned}
\quad\quad
\begin{aligned}
\begin{tikzpicture}
\path [use as bounding box] (-0.25,-0.75) rectangle +(4.5,5);
\draw [surface] (-\xoff,\yoff) to [out=right, in=145] (1,0) to (2,-0) to [out=30, in=left] (3,\yoff) to [out=25, in=left] (4-\xoff,2*\yoff) to (4-\xoff,2) to [out=up, in=up, looseness=0.5] (-\xoff,2) to (-\xoff,\yoff);
\node [label, below] at (-\xoff,\yoff) {$B$};
\draw [black] (4-\xoff,2*\yoff) to (4-\xoff,2) to [out=up, in=up, looseness=0.5] (-\xoff,2) to (-\xoff,\yoff);
\draw [surface] (\xoff,-\yoff) to [out=right, in=-150] (1,0) to [out=up, in=down] (2,1) to [out=up, in=left] (2.5,1.5) to [out=right, in=up] (3,1) to (3,\yoff) to [out=-35, in=left] (4+\xoff,0) to (4+\xoff,3.5) to [out=up, in=up, looseness=0.5] (\xoff,3.5) to (\xoff,-\yoff);
\draw [basic] (1,0) to [out=up, in=down] (2,1) to [out=up, in=left] (2.5,1.5);
\draw [adjoint] (2.5,1.5) to [out=right, in=up] (3,1) to (3,\yoff);
\draw [black] (4+\xoff,0) to (4+\xoff,3.5) to [out=up, in=up, looseness=0.5] (\xoff,3.5) to (\xoff,-\yoff);
\node [label, below, red] at (3,\yoff) {$\otimes$};
\draw [surface] (2,0) to [out=-35, in=left] (4+3*\xoff,-2*\yoff) to [out=up, in=-15, out looseness=1.8] (1,3.2) to (1,1) to [out=down, in=up] (2,0.0) to (2,0);
\draw [bare] (4+3*\xoff,-2*\yoff) to [out=up, in=-15, out looseness=1.8] (1,3.2);
\draw [basic] (4+3*\xoff,-2*\yoff) to [out=up, in=-15, out looseness=1.8] (1,3.2) to (1,1) to [out=down, in=up] (2,0.0) to (2,0);
\node [label, below] at (\xoff,-\yoff) {$A$};
\node [label, below] at (4+\xoff,0) {$A$};
\node [label, below] at (4+3*\xoff,-2*\yoff) {$\bot$};
\node [label, below] at (4-\xoff,2*\yoff) {$B$};
\end{tikzpicture}
\end{aligned}
\quad\quad
\begin{aligned}
\begin{tikzpicture}
\path [use as bounding box] (-0.25,-0.75) rectangle +(4.5,5);
\node [label, below] at (-\xoff,\yoff) {$B$};
\draw [black, thick] (3,1) to [out=60, in=down] (4-\xoff,2) to [out=up, in=up, looseness=0.5] (-\xoff,2) to (-\xoff,\yoff);
\draw [black, thick] (3,1) to [out=140, in=down] (4+\xoff,3.5) to [out=up, in=up, looseness=0.5] (\xoff,3.5) to (\xoff,-\yoff);
\draw [black, thick] (3,\yoff) to (3,1) node [circle, draw=black, text=black, fill=white, thick, inner sep=0.5pt] {$\otimes$};
\node [label, below] at (3,\yoff) {$A\!\otimes \!B$};
\draw [black, thick, densely dotted] (1,3.2) to (\xoff,3.5);
\draw [black, thick] (4+3*\xoff,-2*\yoff) to [out=up, in=-15, out looseness=1.8] (1,3.2) node [circle, draw=black, text=black, fill=white, thick, inner sep=0.5pt] {$\bot$};
\node [label, below] at (\xoff,-\yoff) {$A$};
\node [label, below] at (4+3*\xoff,-2*\yoff) {$\bot$};
\end{tikzpicture}
\end{aligned}
\hspace{-5cm}
$$
\vspace{-10pt}
\caption{\label{fig:comparison}A deduction in the sequent calculus, and its surface calculus and proof net representations.}
\end{figure}
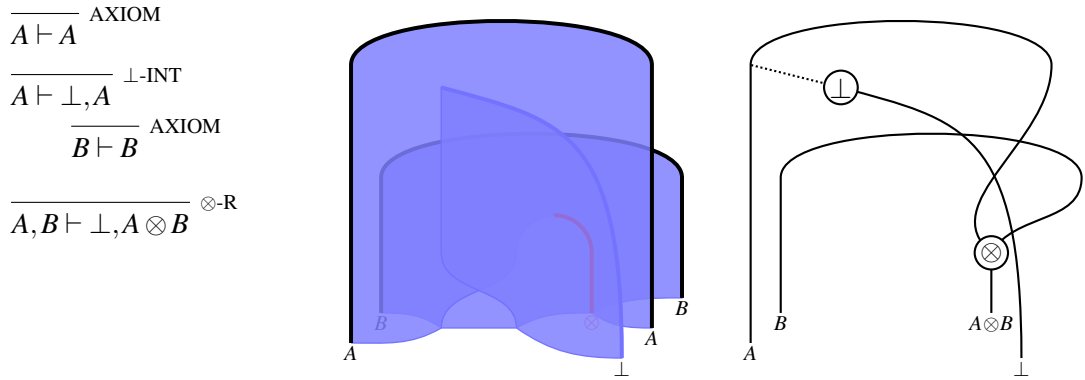%
Choosing $X=\bot$, then $q_A$ and $p_A$ are the isomorphisms $\ls(A^*) \simeq A \simeq (\ls A)^*$. We give the surfaces for the clockwise and anticlockwise paths of \eqref{eq:tripleunit}:
\input{tripledual.extra}

\noindent
We conclude that the proofs are equivalent by a \textit{single} application of the \textbf{Coherence} rule. Contrast this with the treatment of Blute et al~\cite[Section~4.2]{blute-coherence} and Hughes~\cite[Example~2]{hughes-simplestar} in terms of proof nets, where the proofs require several rewiring steps.
\end{example}

\begin{example}
[Equivalence of two proofs with units]
The example is given in \autoref{figure:proofswithunits}. We present two distinct sequent proofs of the tautology $A, B \vdash \bot \parr (A \otimes B)$, along with their corresponding surface proofs. The heights are aligned to help understand how the surface proofs have been constructed. We make use of the $\otimes$-R rule derived in the previous example.

It can be seen by inspection that the surface proofs are equivalent, as follows. Starting with the surface on the left, we allow the $\bot$-introduction vertex to move up and to the left; this is an application of \textbf{Coherence} and \textbf{Locality}. We also allow the $B$-introduction vertex at the top of the diagram to move down, behind both the $A$-introduction and $\bot$-introduction vertices; this is an application of \textbf{Isotopy}.
\end{example}

\begin{example}[Proof net projection] We argued in the introduction that proof nets can be considered the 2d `shadow' of the full 3d geometry, with the correctness criterion and thinning links arising to compensate for the fact that this `shadow' has lost some essential geometrical data. We illustrate this in \autoref{fig:comparison}, which gives a sequent proof alongside its surface diagram and proof net representations. In particular, in the proof-net picture, the bottom attachment must be `hooked' onto a solid wire, indicated by the dotted line, a feature which is absent from the 3d image, which contains connectivity data directly as part of the 3d geometry.
\end{example}

\small
\bibliographystyle{plainurl}
\bibliography{references}

\end{document}